\documentclass[aps,pre,showpacs,twocolumn,floatfix]{revtex4}
\usepackage{graphicx,amsmath,amsfonts,amssymb,color}
\usepackage{epstopdf}
\DeclareGraphicsExtensions{.pdf,.eps,.png,.jpg,.mps} 

\begin{document}

\title{Synchronization in Phase-Coupled Kuramoto Oscillator Networks with Axonal Delay and Synaptic Plasticity} 
\author{L. Timms$^{1}$ and L. Q. English$^{1}$ }
\affiliation{$^{1}$Department of Physics and Astronomy  \\ 
Dickinson College, Carlisle, Pennsylvania, 17013, USA}
\date{\today}

\begin{abstract}
We explore both analytically and numerically an ensemble of coupled phase-oscillators governed by a Kuramoto-type system of differential equations. However, we have included the effects of time-delay (due to finite signal-propagation speeds) and network plasticity (via dynamic coupling constants) inspired by the Hebbian learning rule in neuroscience. When time-delay and learning effects combine, novel synchronization phenomena are observed. We investigate the formation of spatio-temporal patterns in both one- and two-dimensional oscillator lattices {with periodic boundary conditions} and comment on the role of dimensionality. 
\end{abstract}


\maketitle

\section{Introduction}
\label{sec:intro}

Just as simple physical oscillators can synchronize when coupled together, neural synchronization on the level of both individual cells and larger brain regions has been observed in a variety of instances. For example, adaptive behavioral processes appear to be linked to collective oscillatory activity in the cortex \cite{Breakspear}, and other research has indicated that neuronal synchronization plays a major role in motor-function, Parkinson's disease, epilepsy, resting state functions, and memory \cite{Breakspear, Hoppensteadt99, Galtier}. 

A neuron is a cell featuring dendrites and an axon that allow it to communicate with a large number of other neurons at varying distances (between cell nuclei). This communication typically proceeds by first integrating inputs coming in at the dendrites; if a threshold is reached, an output in the form of action potentials is generated that travels down the axon. The integrated input signals can be excitatory (push the receiving cell towards its threshold) or inhibitory (pull the receiving cell away from its threshold).
Additionally, certain neurons behave ``tonically" and can create action potentials periodically by self-stimulation. Due to this periodicity, such cells can be modeled individually as oscillators. Additionally, neurons may be arranged in networks like Central Pattern Generators in which the entire local network acts as an oscillator \cite{Rand, Varkonyi, Ijspeert}. 

The Kuramoto model represents a famously tractable, but also fairly generic, model of mass-synchronization in biological systems \cite{kura, strogatzD}. It has been applied to a plethora of phenomena  \cite{Acebron, StrogatzInsight, Wiesenfeld, Solomon, Mertens}, and is sometimes proposed in connection with neural network dynamics. Yet, due to the complexity of interactions in neural networks there has been relatively few direct application of the Kuramoto model to neuro-biological data \cite{Perez}. To perhaps begin to remedy this situation, we propose a combination of two extensions or modifications to the Kuramoto model that are very important in the neuroscience context: the inclusion of spatial embedding through time-delays and the inclusion of variable synaptic strength through dynamically changing coupling. 

These two modifications together are most likely indispensable if this model is to be successful in describing many aspects of neural processes. First, time-delays are essential because neuronal axons have finite signal transmission speeds \cite{KoandErmentrout}. The fact that signals often take physiologically significant time to reach their destinations is fundamental to the design of neural networks, and the lack of time-delays in the original Kuramoto model has been identified as a significant obstacle \cite{Acebron, Breakspear, Perez}. Second, the brain must adapt in some way to accommodate new functions, memories and skill. The adaptation of individual neural networks occurs through synaptic modification according to mechanisms on the scale of single neurons. These modifications are theorized to be the cellular basis of learning and long-term memory \cite{Acebron, Lars, Soltoggio} and may be required for the creation of neurocomputers \cite{Hoppensteadt99}.

Neurophysiology tells us that certain synapses can be strengthened or weakened as a function of how often two neurons fire together. The precise way in which this concept should be represented mathematically is still a matter of debate in the literature. One of the simplest ways of proceeding is the delta-formulation of the Hebbian Learning Rule \cite{Galtier, Papakostas}, and we follow this approach also taken elsewhere \cite{Seliger, Lars}. 

In this study, we examine both one- and two-dimensional globally-communicating oscillator arrays with periodic boundary conditions, where time-delays are computed based on the Euclidean distance between nodes, consistent with previous approaches \cite{Zanette, Yeung, Perez}. However, we also include the effects of Hebbian learning and examine the novel interplay of these two additions to the Kuramoto model. In the one-dimensional setting, the fast learning case is perhaps most interesting, as it exhibits new modes of spatio-temporal order. Here, we derive self-consistency equations that yield the synchronization frequency as a function of signal speed. A comparison of these analytical results to numerical simulations shows close agreement. In the two-dimensional lattice, spatio-temporal patterns other than plane-waves can arise; for faster learning, for instance, spiral-like patterns appear to dominate. 


\section{The Model}
\label{sec:model}
The well-known Kuramoto model serves as a convenient starting point; recall that it describes the evolution of a set of $N$ oscillators whose states are described by phase alone, $\phi_i (t)$. The actual time evolution of these phases is governed by a system of coupled differential equations which include phase-interaction terms between the oscillators. This system of equations is given by
\begin{equation}
\dot{\phi_i} (t)= \omega_i  + \frac{K}{N}  \sum\limits_{j=1}^N \sin(\phi_j(t)-\phi_i(t)).
\label{eq:Basic}
\end{equation}
Here, $\dot{\phi_i}$ is the time derivative of the phase, $\omega_i $ is the natural frequency of the $i^{th}$ oscillator, $K$ is the coupling strength between all oscillator pairs, and $N$ is the total number of oscillators. As was famously proved by Kuramoto analytically, the sine-coupling term drives the oscillators towards synchronization, but only if the coupling constant $K$ exceeds a critical value of $K_c=\sqrt{8/\pi}\sigma$, where $\sigma$ represents the standard deviation in the $\omega_i$ distribution \cite{kura, strogatzD, lars05}.

To adapt this to our model, we first relax the condition that the coupling strengths between all pairs of oscillators be equal. To this end, we define $K_{ij}$ to be the connection strength of oscillator $i$ to oscillator $j$; in general this matrix will not be symmetric (especially with the inclusion of time delays). Second, as is the case for all physical systems, we require the transmission speed of information to be finite, forcing the inclusion of time-delays \cite{Yeung, Zanette, Jeong, Ko}. In the neurophysiological context, we represent a difference in relative spike timing by a difference in phase; the reference the neuron is using is the time at which the incloming signal is actually received, and so accounting for communication delay is necessary. The easiest way to include a time-delay is to modify Eq.~(\ref{eq:Basic}) to
\begin{equation}
\dot{\phi_i} (t)= \omega_i  + \frac{1}{N}  \sum\limits_{j=1}^N K_{ij} \sin(\phi_j(t-\tau_{ij})-\phi_i(t)).
\label{eq:timedelay}
\end{equation}
Here, $ \tau_{ij} \equiv d_{ij}/v $ where $d_{ij}$ is the distance from oscillator $i$ to oscillator $j$ and $v$ is the velocity of the signal. In this formulation the relative spatial positions of the oscillators become important. Yet the coupling strengths may be independent of any spatial arrangement, and this represents a break from many previous studies \cite{Acebron}. Note that a change in $\tau_{ij}$ can correspond to changing the distance between components or changing the velocity of the signal. It is also worth highlighting that ours is a more direct approach than the one taken in many delay-focused studies in which the phase velocity is assumed to be constant, and the phases are then simply adjusted by an appropriate factor \cite{Breakspear}, $\sin(\phi_j-\phi_i-\alpha_{ij})$. As we will see, the assumption of a fixed phase velocity (independent of other parameters) is generally not justified.

Third, we must allow network plasticity in order to examine the role of ``learning'' and generally investigate the mutual interaction between oscillator dynamics and network dynamics.
We implement a version of the Hebbian learning rule into the model that increases connection-strengths between those oscillators whose phases are close in value \cite{Seliger, Lars, Aoki}. Thus, we define a second set of differential equations,
\begin{equation}
\dot{K}_{ij} (t)= \epsilon[\alpha \cos(\phi_i (t) - \phi_j (t- \tau_{ij})) - K_{ij}].
\label{eq:coupling}
\end{equation}
Here, $\epsilon$ is a constant that allows us to arbitrarily adjust how ``dynamic" the coupling is, i.e., how fast changes in coupling strengths can occur, and the cosine-function forces the coupling strengths to increase when the difference in phase approaches zero and decrease when it approaches $\pi$ (exactly out-of-phase). Additionally,  the parameter $\alpha $ in Eq.~\eqref{eq:coupling} provides the fixed-point value which $K_{ij}$ would approach if the $ij$-pair were perfectly phase-matched. Thus, conceptually $\alpha$ and $-\alpha$  are the maximum  and minimum values that the coupling is allowed to attain, respectively. 

Equation \eqref{eq:coupling} differs from the model used in Ref.\cite{Seliger, Lars} only by the inclusion of time delay in the phase used. This inclusion is justified, as the Hebbian learning rule is concerned with relative timing of action potentials of the pre- and post-synaptic cells (the cell before the connection, or ``synapse," and the cell after it, respectively). Hence, we choose to include $\tau_{ij}$  because the interaction we wish to model only occurs when an action potential actually reaches the synapse. Note that we are using the phrase ``Hebbian Learning'' somewhat liberally here; this is not the traditional rule but rather its Delta Formulation. Furthermore, we are making another simplification here by not restricting the coupling to positive or negative values. This means the interaction between oscillator pairs can easily switch from excitatory to inhibitory.


Having arrived at our mathematical model, the larger question we would like to address is this: can the inclusion of time delay modify in nontrivial ways the spatio-temporal ordering observed in Ref. \cite{Lars} for instantaneous neuronal communication; as a new parameter of the system, what detailed role does the communication speed play in bringing about new dynamical features exhibited by the system. Recall that the inclusion of network plasticity (in the form of Hebbian learning) resulted in the emergence of two stable clusters ($\pi$ out of phase with one another) \cite{Lars}, and this was seen to occur because oscillators in one cluster were positively coupled to each other but negatively coupled to those in the opposing cluster. How does time delay affect those clusters? As we will see, the interplay between two cluster synchronzation and time delay will yield novel patterns of synchronization.    

\section{Analytical Results}
\label{sec:analytical}
Intriguingly, in the dynamical system given by Eqs.~\eqref{eq:timedelay} and \eqref{eq:coupling} some global properties of the emergent synchronization patterns can be predicted analytically. Much of this work follows a procedure outlined in Ref.\cite{Zanette} but novelly incorporates network plasticity.  

To set the stage, let us 
define synchronization behavior to be organizations where all the oscillators move with the same frequency, $\dot{\phi_i}=\Omega$, at long times ($t\rightarrow\infty$). This implies that synchronization behavior is not defined by all oscillators converging in phase but rather in frequency. Certainly, they may all converge in phase in special circumstances (as we shall see) but in general, we can give each oscillator an arbitrary but constant phase-offset, $\psi_i$. This definition of sync and the use of phase offset will allow us to consistently {capture} the spatial patterns which may form in response to time-delays. The synchronized state is then governed by $\phi_i(t)=\Omega t + \psi_i$. Thus,
  
\begin{eqnarray}
\phi_j(t-\tau_{ij})-\phi_i(t)&=&\Omega \cdot (t-\tau_{ij})+\psi_j-(\Omega t + \psi_i) \nonumber \\
&=&-\Omega \tau_{ij}+\psi_j-\psi_i,
\label{eq:Omega}
\end{eqnarray}
which we can now insert into Eq.~\eqref{eq:timedelay} to obtain,
\begin{equation}
\Omega= \omega_i - \frac{1}{N}  \sum\limits_{j=1}^N K_{ij}\sin(\Omega\tau_{ij}+\psi_i-\psi_j).
\label{eq:Omega2}
\end{equation}
Previous work has shown that in the fast learning regime, $\epsilon>\epsilon_c$, Eq.~\eqref{eq:coupling} leads to coupling strengths governed by $K_{ij}(t)= K^*_{ij}=\alpha \cos(\phi_i-\phi_j)$, which means that the coupling strengths adiabatically follow their fixed points in Eq.~\eqref{eq:coupling} \cite{Lars}. Adapting this result to our model we can write, 
\begin{equation}
 K_{ij}(t)=\alpha \cos(\Omega\tau_{ij}+\psi_i-\psi_j).
\label{eq:Kstable}
\end{equation}
Combining the last two equations, we arrive at,
\begin{equation}
\Omega= \omega_i - \frac{\alpha}{2N}  \sum\limits_{j=1}^N \sin2(\Omega\tau_{ij}+\psi_i-\psi_j),
\label{eq:OmegaFast}
\end{equation}
via a trigonometric identity.

This equation's validity is dependent on both the system's frequencies converging and the parameter $\epsilon$ being sufficiently high. Notice that Eq.~\eqref{eq:OmegaFast} is not a differential equation but a time-independent transcendental equation that can be solved for $\Omega$ by numerical means, once the spatial dependences of $\tau_{ij}$ and $\psi_j-\psi_i$ have been specified. 

In order to proceed, we now need to stipulate a concrete network structure. We begin with the one-dimensional (1D) line, for it is the simplest arrangement in which time-delays have nontrivial dynamical implications. Again for simplicity, we impose periodic boundary conditions; then the distance is given by $ d_{ij} =  L/N  \min({|i-j|, N-|i-j|})$, where $L$ is the linear size of the array. Next, we define a new time-delay parameter $T$, such that $T \equiv L/v$. It follows that, 
\begin{equation}
\tau_{ij}=\frac{T}{N}\min({|i-j|, N-|i-j|}),
\label{eq:tau}
\end{equation}
therefore we always chose the shorter way around the circle between two given oscillators \cite{Zanette}. This implies there is only one connection between each pair; the signal cannot travel both ways around the ring to get to the same node at different times. Changing the value of the parameter, $T$, allows us to model a change in either the speed of the signal (inversely related to $T$) or the overall spatial scale of the system. 

Investigation of phase-coupled oscillator systems with time-delay has previously lead to the discovery of different coherent-wave modes of synchronization \cite{Zanette, Jeong}. In 1D arrays, these modes are described by a static phase increment from one oscillator to the next. Further, in order to satisfy periodic boundary conditions, the total phase advance over the system length should be a multiple of $2\pi$. Without loss of generality, if we take the $i$-th oscillator to be our starting point, $i=1$ and $\psi_1 =0 $, then there are multiple standing-wave modes, indexed by integer $m$, satisfying the periodic boundary condition. Moreover, as mentioned earlier, this dynamic-coupling allows two clusters in anti-phase to arise \cite{Lars}. Hence the modes are now given by,
\begin{eqnarray}
\psi_j^{(a)}&=& 2\pi \frac{m}{N} (j-1) \nonumber \\
\psi_j^{(b)}&=& \psi_j^{(a)}+\pi,
\label{eq:psi}
\end{eqnarray}
where the superscript denotes the cluster. The way to interpret Eq.\eqref{eq:psi} is that the phase of oscillator $j$ can either be given by the first equation or the second, depending on the cluster with which this oscillator is associated. In other words, if we plot all oscillator phases against the spatial index $j$, we see points that fall on the two sloped lines given by Eq.\eqref{eq:psi} in random alteration.

Interestingly, unlike the one-cluster situation where all modes are given by one of the lines of Eq.~\eqref{eq:psi} \cite{Zanette}, we now get new ways of satisfying the periodic boundary conditions.
To see how these additional modes come about, imagine that oscillator 1 happens to be in cluster (a) and let's advance along the line until we get to oscillator $N$. This oscillator could conceivably be part of cluster (b), and therefore only a phase difference of $\pi$ would have had to accumulate. Now {turn to} oscillator 2, which could start out in cluster (b). As we advance across the line and end at oscillator 1, in cluster (a), we have again accumulated a total phase of { $\pi$}. Thus, we get additional ``inter-connected cluster'' modes of the form,
\begin{eqnarray}
\psi_j^{(a)}&=& (2\pi m-\pi)\frac{(j-1)}{N} = 2 \pi \frac{m-1/2}{N} (j-1)\nonumber \\
\psi_j^{(b)}&=& \psi_j^{(a)}+\pi
\label{eq:psi2}
\end{eqnarray}
These mode accomodate the time delay by advancing the phase along one direction of the ring only by $\pi$. The reason this does not cause a phase discontinuity somewhere along the ring is due to the presence of two clusters. Thus, such modes are not possible in the separate time-delayed models or the Hebbian learning models, they fundamentally require both.  

\begin{figure}
\includegraphics[width=2.65in]{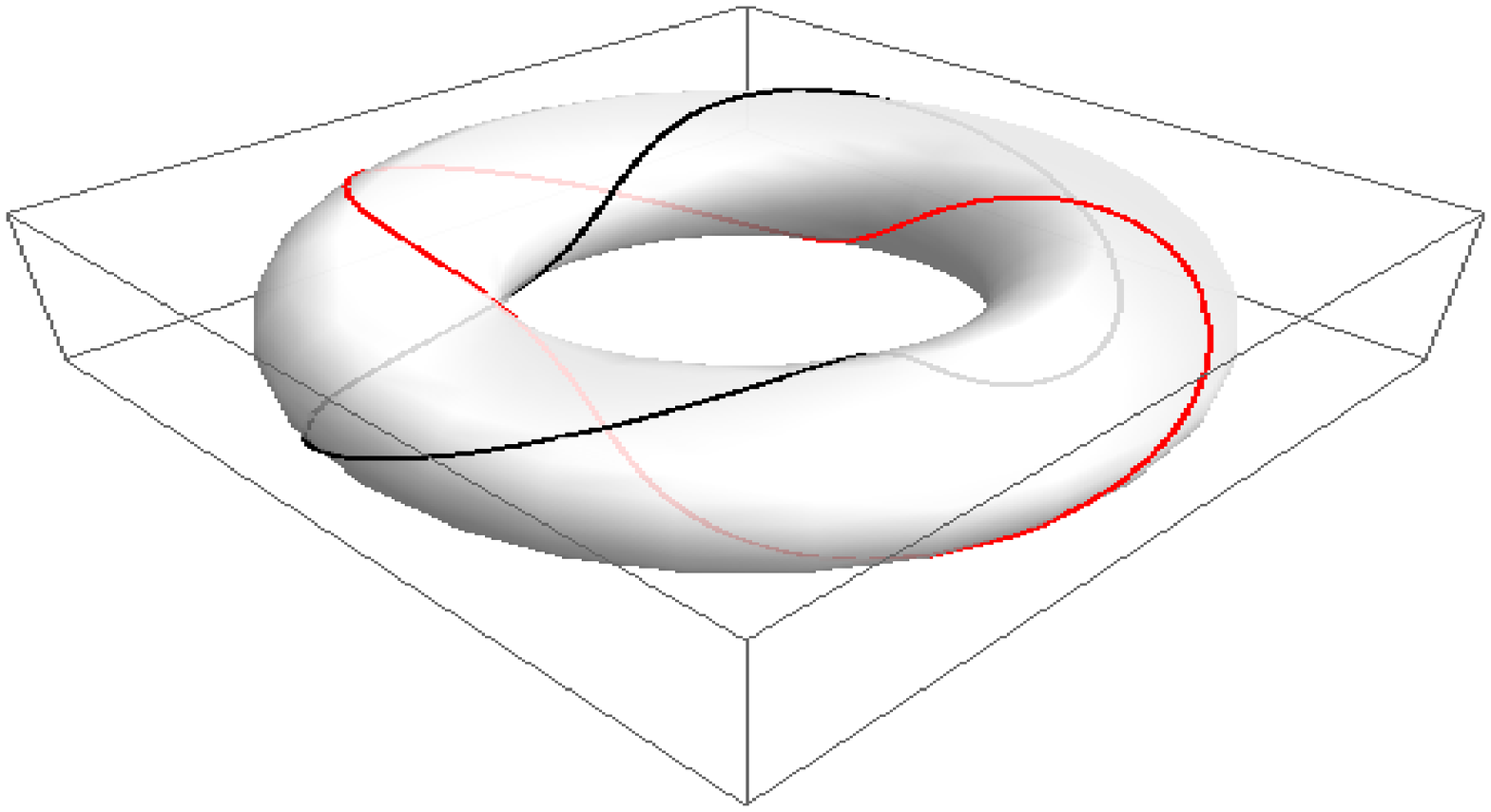}
\includegraphics[width=2.65in]{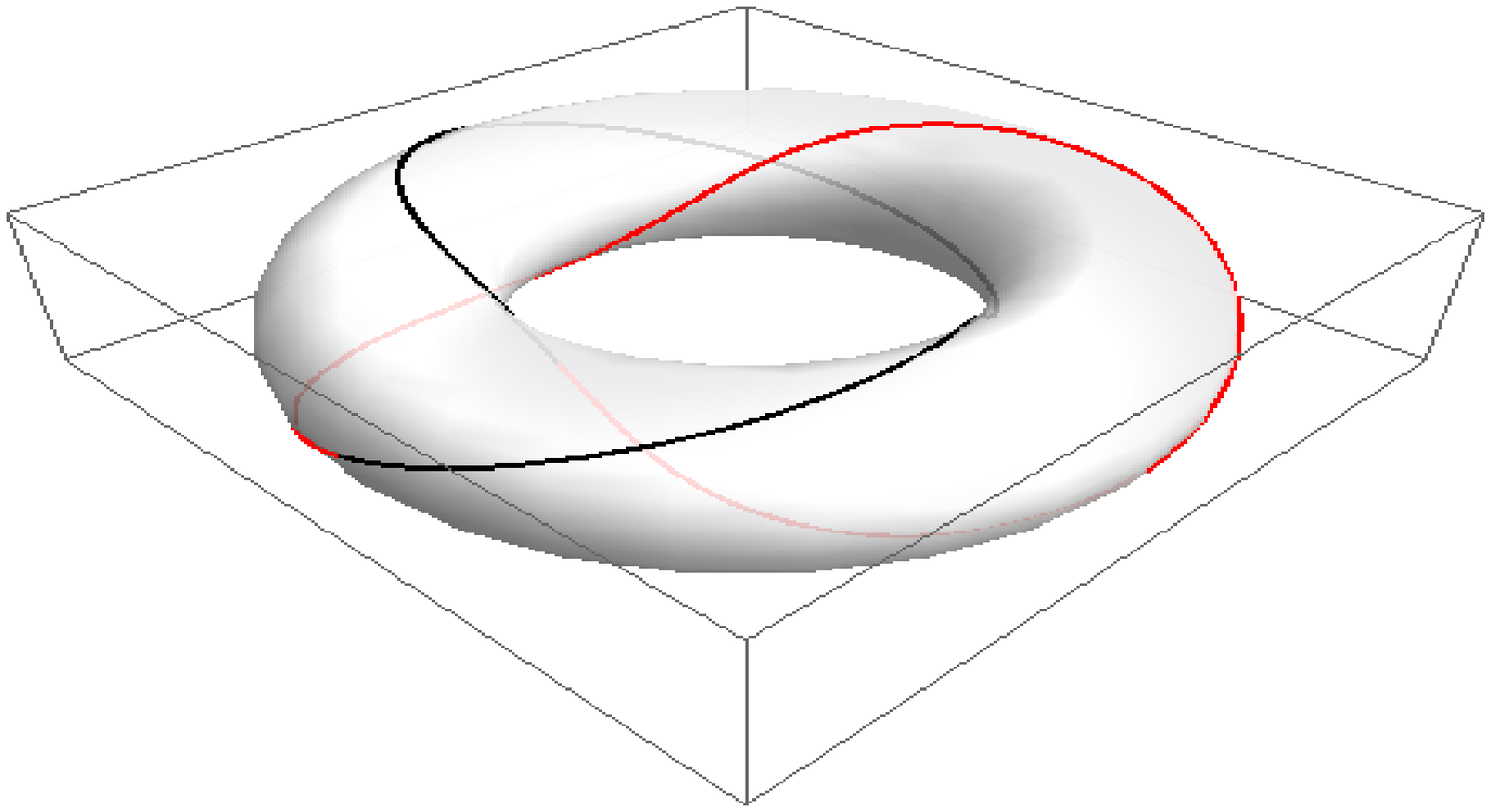}
\caption{Visual representation of the allowed modes for stable two-cluster arrangements in the case of $m=2$. Note that the lower panel shows the novel mode with two clusters (black and red) inter-connecting. In the rest of the paper, these interconnecting modes will be represented by half-integer $m$ (see text). Thus, in this representation, the upper and lower panels show $m=2$ and $m=1.5$, respectively.}
\label{Torus}
\end{figure}

One way to visualize these modes is on a torus, where the circular direction around the torus center represents the spatial oscillator ring, and the torus tube represents the oscillators' phase from zero to $2\pi$ as if on a unit circle. This is shown in Fig.\ref{Torus}, where the $m=2$ mode is selected for the two cases. 

Equations \eqref{eq:psi2} for the interconnecting 2-cluster modes can be described by Eq.~\eqref{eq:psi} if we allow $m$ to take on half-integer values in addition to integer values. We will follow this convention in the rest of the paper. 
We proceed by inserting into Eq.~\eqref{eq:OmegaFast} the expression for $\tau_{ij}$, given by Eq.~\eqref{eq:tau}, as well as the mode equation, given by Eq.~\eqref{eq:psi}, and arrive at a convenient self-consistency condition on $\Omega$,  
\begin{eqnarray}
& \Omega= \omega_1 - \frac{\alpha}{2N} \sum\limits_{j=1}^N \nonumber \\
 &  \sin 2\left[ \Omega \frac{T}{N}\min(j-1, N-j+1)
-2 \pi \frac{m}{N} (j-1) \right].
\label{eq:Omega1D}
\end{eqnarray}
Finally, we make the simplifying assumption that $\omega_1=\omega_i=1$, although as we will see later this assumption of no spread in intrinsic frequencies is not crucial. We can then employ numerical methods \cite{mathematica} to find $\Omega$ for various choices of $T$ and $m$. This equation is analogous to the self-consistency equation in Ref.\cite{Zanette} which in our formulation would be obtained in the special case of $\epsilon=0$, i.e., in the case of a static network with no learning.

\begin{figure}
\centering
\includegraphics[width=3.25in]{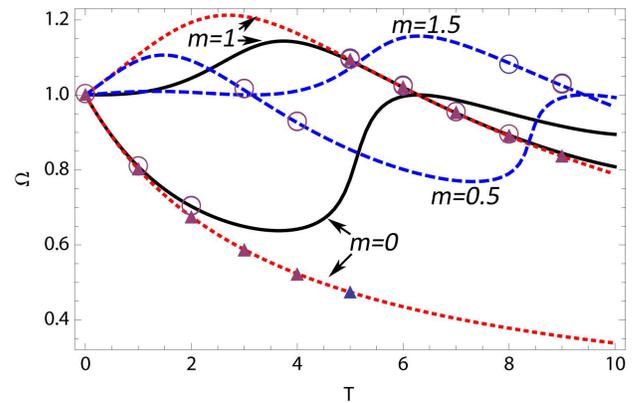}
\caption{Lines correspond to solutions of the self-consistency equations, such as Eq.~\eqref{eq:OmegaFast}: solid and dashed lines depict the fast learning case for integer and half-integer modes, respectively; the dotted lines are for the no-learning case where only one cluster (and therefore integer $m$) is possible.}
\label{OmegaTm}
\end{figure}

The results are summarized in Fig.~\ref{OmegaTm}. Here the lines are the theoretical prediction of Eq.~\eqref{eq:Omega1D} (and its version for $\epsilon=0$), for $m=0, 0.5, 1$ and $1.5$, as functions of the delay-parameter $T$. The dotted lines are the results for the no-learning case ($\epsilon=0$), the solid lines represent fast learning ($\epsilon=0.1$) but for integer $m$, and finally the dashed lines depict the fast-learning case for half-integer $m$. The markers depict the observed results from numerical simulations (see also next section). We see excellent agreement with numerical results in all instances. Note that although the analytical approach yields the correct synchronization frequencies $\Omega$, it cannot tell us when transition between the different allowed modes will occur. This can only be ascertained numerically, as we will further discuss in the next section.
We should note that in two-dimensional arrays, which we have also studied, many of the previous equations remain unchanged, but the $\tau_{ij}$ expression has a different form. Let's define $X$ to be the number of oscillators along the $x$ coordinate and $Y$ to be the number along the $y$ coordinate, and so $N = XY$. If we pick two arbitrary oscillators lying on a two-dimensional $X \times Y$ lattice, $\vec{r}_1=(i,j)$ and $\vec{r}_2=(k,l)$, then the time-delay between them can be expressed as,
\begin{eqnarray}
\label{eq:tau2D}
&\tau_{ij,kl}=\frac{T}{\sqrt{XY}} \times \\
&\sqrt{(\min({|i-k|, X-|i-k|})^2 + (\min({|j-l|, Y-|j-l|})^2}. \nonumber
\end{eqnarray}

\section{Numerical Results}
\label{sec:numerical}

\subsection{One-Dimensional Array}
\label{subset:1D}
To solve the dynamical system, given by Eqs.~(\ref{eq:timedelay}) \& (\ref{eq:coupling}), numerically, we choose $N=100$ and initialize the coupling strengths at $K_{ij}(0)=1$ for all pairs and parameter values. The spread in natural frequencies was obtained by random assignment according to a Gaussian probability density centered on 1 with a width of 0.1. Initial conditions for the oscillators' phase were given by noise functions such that they were spread out in phase at random, with no spatial structure. An Euler-method was used to evolve the system forward in time at a time-step of $dt=0.01$ which was shown to be appropriate by a previous study \cite{Zanette}. The time delays were implemented by storing all $\phi$'s at each time-step in a massive table and then calling the $\phi$ value directly for each time, $t-\tau_{ij}$. To allow the program to have values to call upon initially, the system was allowed to evolve according to only the natural frequencies for 1000 time steps before the interaction was turned on. Additionally, for all data shown here, $\alpha=1$.

The numerical results displayed in Fig.\ref{OmegaTm} show that as the time-delay parameter, $T$, is increased (or equivalently as the propagation speed decreased), sharp transitions occur between the actual modes realized by the synchronizing oscillators; this is true in both no-learning and fast-learning cases. In the fast-learning case, the $m=0$ mode of full synchronization is established at low $T$ values, although the synchronization frequency drops. Then, somewhere within the interval  $2\leq T \leq 3$, a transition to the $m=0.5$ mode occurs, which is followed by another transition to the $m=1$ mode, and finally for $T \geq 8$, we end up in mode $m=1.5$. What is interesting is that the interconnected (half-integer) modes form a bridge between the non-connected (integer) modes at large enough $\epsilon$. For small $\epsilon$ these bridges cannot happen because in the case of one cluster, the half-integer modes are forbidden.

\begin{figure}
\centering
\includegraphics[width=3.25in]{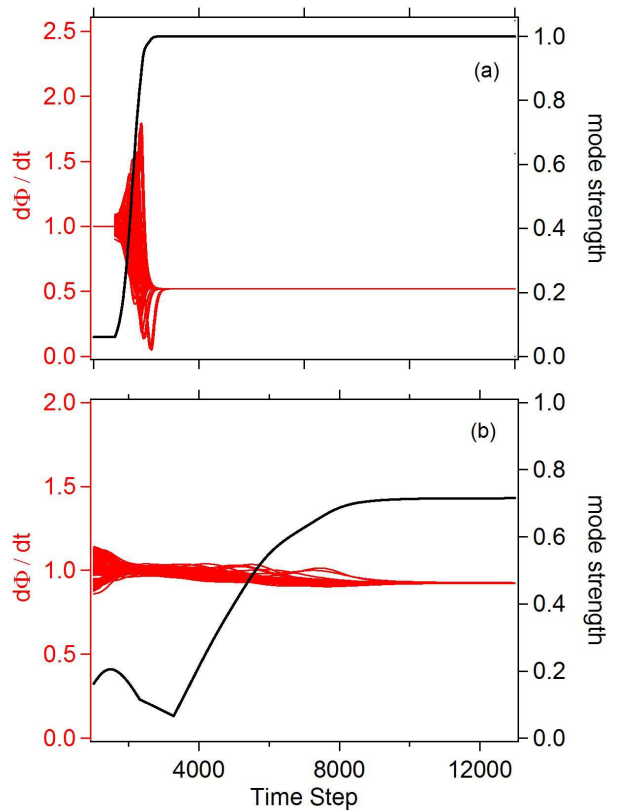}
\caption{The time-evolutions of all oscillators' phase velocities, $\dot{\phi}_i$ (left axis), compared against the order parameter measuring the strength of a particular mode of synchronization (see text for definition). $T$ is set to 4. (a) $\epsilon=0$ and (b) $\epsilon=0.1$.}
\label{op}
\end{figure}

Let us now examine the detailed approach to the synchronized state and, for concreteness, isolate from Fig.\ref{OmegaTm} the two datapoints at $T=4$ corresponding to no-learning ($\epsilon=0$) and fast learning ($\epsilon=0.1$), respectively. Panel (a) of Fig.\ref{op} depicts the former case while (b) depicts the latter. The cluster of traces 
show the time-evolution of the phase-velocities, $\dot{\phi}_i$, of all $N=100$ oscillators. We see that initially (when the model kicks in at t=1000), there is a spread in phase-velocities and oscillators arrange themselves in phase in an attempt to find a synchronized state. In the no-learning case of Fig.\ref{op}(a), for instance, the phase-velocities begin to converge after around 2500 time steps, and after t=3000 they have fully synchronized at a global frequency, $\Omega$, of just about half the intrinsic frequencies of $\omega_i=1$. In this case, the synchronized state corresponds to mode $m=0$ (see also Fig.\ref{OmegaTm}) where the oscillators actually fall into a single common phase as well (for the case without spread in $\omega_i$). The only effect of time-delay, here, is to drastically lower the synchronization frequency. 

What happens to this state when learning is turned on? In Fig.\ref{op}(b), we see that the convergence of the phase-velocities takes longer (about 10,000 time steps) to achieve, and that the global frequency lies much closer to the intrinsic frequencies centered on 1. This is possible because the oscillator phases align themselves with mode $m=0.5$, characterized by a steady phase advance from one oscillator to the next. In this mode, what is therefore possible is that within all oscillator-pairs, one partner ``sees'' the other as nearly in phase with itself accounting for the time delay, and no drastic adjustment in $\Omega$ is needed.

The single line in Fig. \ref{op} depicts the order parameter measuring the strength of the relevant mode of synchronization. In panel (a), since $m=0$, this is just the usual order parameter, $r$, as given by
\begin{equation}
r e^{\imath \psi}=\frac{1}{N}\sum_{j=1}^{N} e^{\imath [\phi_j \pm 2\pi m (j-1)/N]}, 
\label{eq:op1}
\end{equation}
where we have corrected for the spatial mode in a way analogous to Eq.~\eqref{eq:psi}(a). In panel (b), we compute the order parameter in the following manner (see also Ref.\cite{Lars}),
\begin{eqnarray}
r' e^{\imath \psi'} &=& \frac{1}{N}\sum_{j=1}^{N} e^{2 \imath [\phi_j \pm 2\pi m (j-1)/N]},\\
r_2^2 &=& |r'-r|^2 ,
\label{eq:op2}
\end{eqnarray}
where $m$ is the mode number for the two-cluster type behavior (following the $m=0,0.5,1,1.5...$ convention) and $r_2$ is the order-parameter. Specifically in the figure we are displaying $m=0.5$, i.e., the first mode in which we have interconnected clusters. Defined in this way, all order parameters measuring the mode strengths range from 0 to 1.
If we compare the order parameter trace to the cluster-evolution of $\dot{\phi}_i$ in Fig.\ref{op}, we notice that the global frequency co-emerges with the mode pattern. We cannot say that one precedes the other, or that one is caused by the other; they always appear together.

\begin{figure}
\includegraphics[width=3.25in]{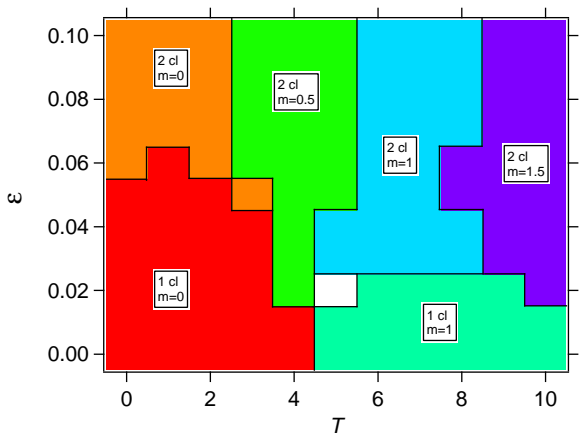}
\caption{Summary of mode synchronization in the parameter space of $T$ and $\epsilon$. Note that for horizontal transitions between two modes, near the transition either of the two modes can be realized depending on initial conditions. Vertically, the transition is somewhat more gradual than indicated here. Nevertheless, the plot gives an overall accurate picture of mode distribution.}
\label{EpsT}
\end{figure}

Figure \ref{EpsT} summarizes the qualitative synchronization behavior in parameter space defined by $\epsilon$ and $T$. For small values of $\epsilon$ (near the bottom of the graph), we can only have integer mode numbers $m$ due to the lack of two-cluster formation. Thus, as the time-delay parameter $T$ is increased, we see a transition from $m=0$ (full synchronicity) to $m=1$. However, when the learning rate $\epsilon$ is above a threshold, two clusters emerge, and this enables half-integer modes. Here, the transition is from $m=0$ to $m=0.5$, and so on, as $T$ is stepped up. Furthermore, what is interesting to note is that the $m=0.5$ mode extends to much lower $\epsilon$-values than would be expected from the situation at $T=0$. For instance, at $T=0$, the transition in $\epsilon$ from one to two clusters happens at around $\epsilon=0.06$. However, at $T=4$, the transition occurs at around $\epsilon = 0.02$. 
The qualitative reason for observing two-cluster formation at much lower $\epsilon$ for larger $T$ is that the oscillator array can ``relieve'' the tension introduced by time-delay by reorganizing into mode $m=0.5$, but this mode is only possible in the two-cluster scenario. 

\begin{figure}
\includegraphics[width=3.25in]{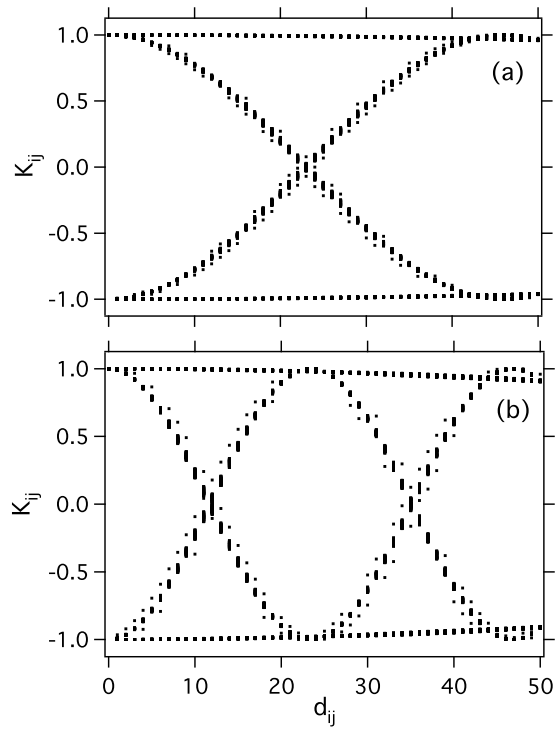}
\caption{The final coupling strengths, $K_{ij}$, as a function of the distance between oscillator pairs, $d_{ij}$ for $\epsilon = 0.1$. (a) $T=4$ and so $m=0.5$, (b) $T=8$ and so $m=1$. In both cases, we have two groups of coupling constants; half of them are $\pm 1$ independent of the distance, and the other half fall on a cosine profile with distance.}
\label{Kvd}
\end{figure}

Finally, before turning to the two-dimensional lattice, let us look at the coupling strengths that are realized for specific modes. Figure \ref{Kvd} plots the coupling strengths after synchronization has been established (for $\epsilon=0.1$) with the modes: $m=0.5$ in Fig.\ref{Kvd}(a), and $m=1$ in Fig.\ref{Kvd}(b). Each coupling strength, $K_{ij}$, is represented by one dot in the figure; the x-axis records the corresponding distance, $d_{ij}$ between the oscillator pair. In both panels, we see that many coupling strengths are simpl{y} $\pm \alpha=\pm 1$, independent of the distance between the oscillators. This is not surprising since the mode is supposed to accommodate the time delay of the signal, and so oscillators will see those oscillators in the direction of wave propagation as in phase or anti-phase with themselves and thus follow $K_{ij}^*=\pm \alpha$. 
However, oscillators in the opposite direction will be seen as progressively further and further out-of-phase. Consequently, we would expect half of the set of $K_{ij}$ to follow $K_{ij}^*=\alpha \cos(\Delta \phi) \neq \pm \alpha$. This is clearly observed in the graph.    
For $m=0.5$, this cosine-contribution connects between groups at $K=1$ and $K=-1$, in contrast to $m=1$, for which the wavelength of the cosine is smaller.

\subsection{Two-Dimensional Array}
\label{subset:2D}
We now turn to the two-dimensional case, where we embed the oscillators in an equally spaced planar array with periodic boundary conditions. This is formally defined in Eq.~\eqref{eq:tau2D}. Prior to presenting our numerical findings we mention a few conceptual implications of the new dimension. Most simply, with the added dimension it is to be expected that more complex patterns of spatio-temporal order are both imaginable and observable \cite{Jeong}. Indeed, previous numerical studies have shown that in the statically coupled case, plane waves, rotating spiral waves and many other complex spatial patterns are possible.
\begin{figure}
\includegraphics[width=3.0in]{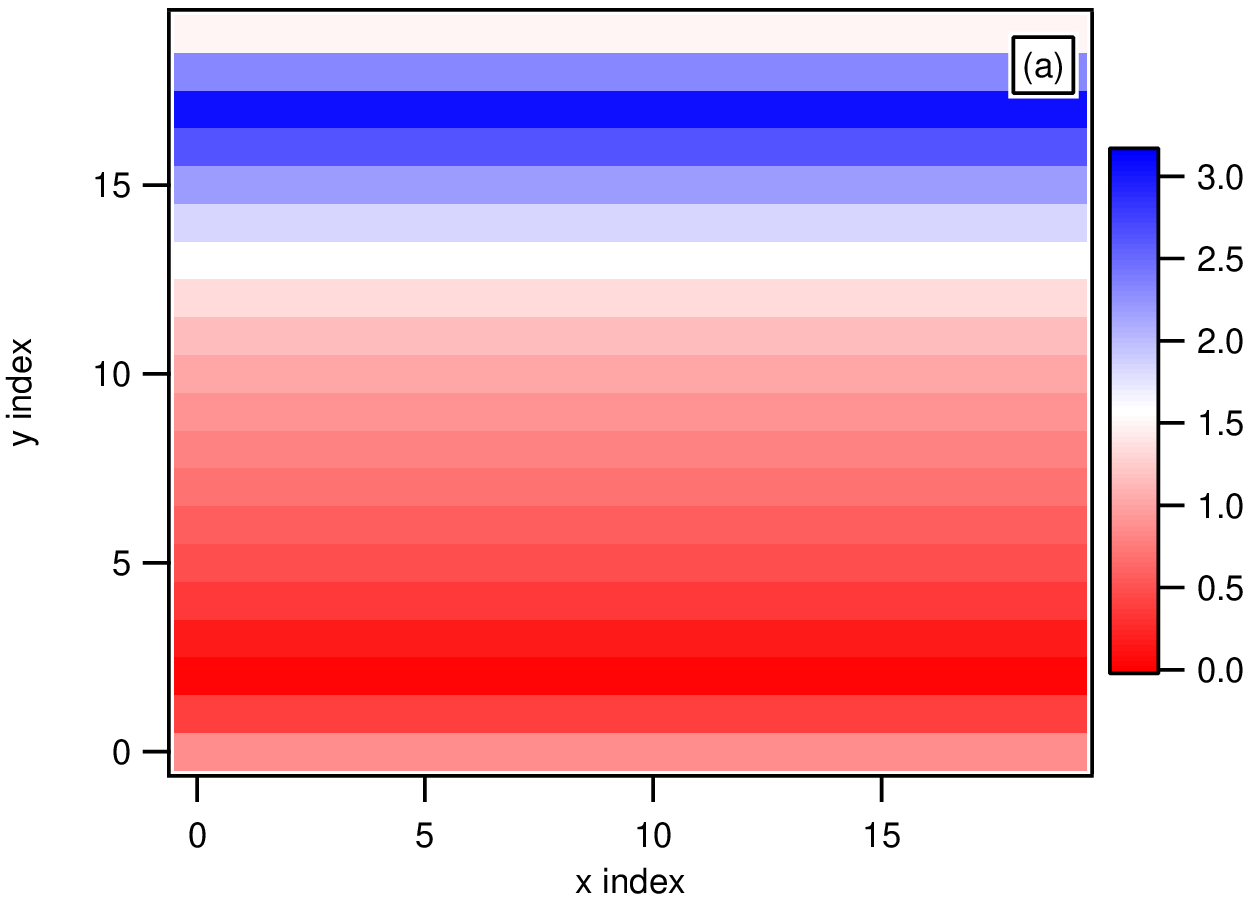}
\includegraphics[width=3.0in]{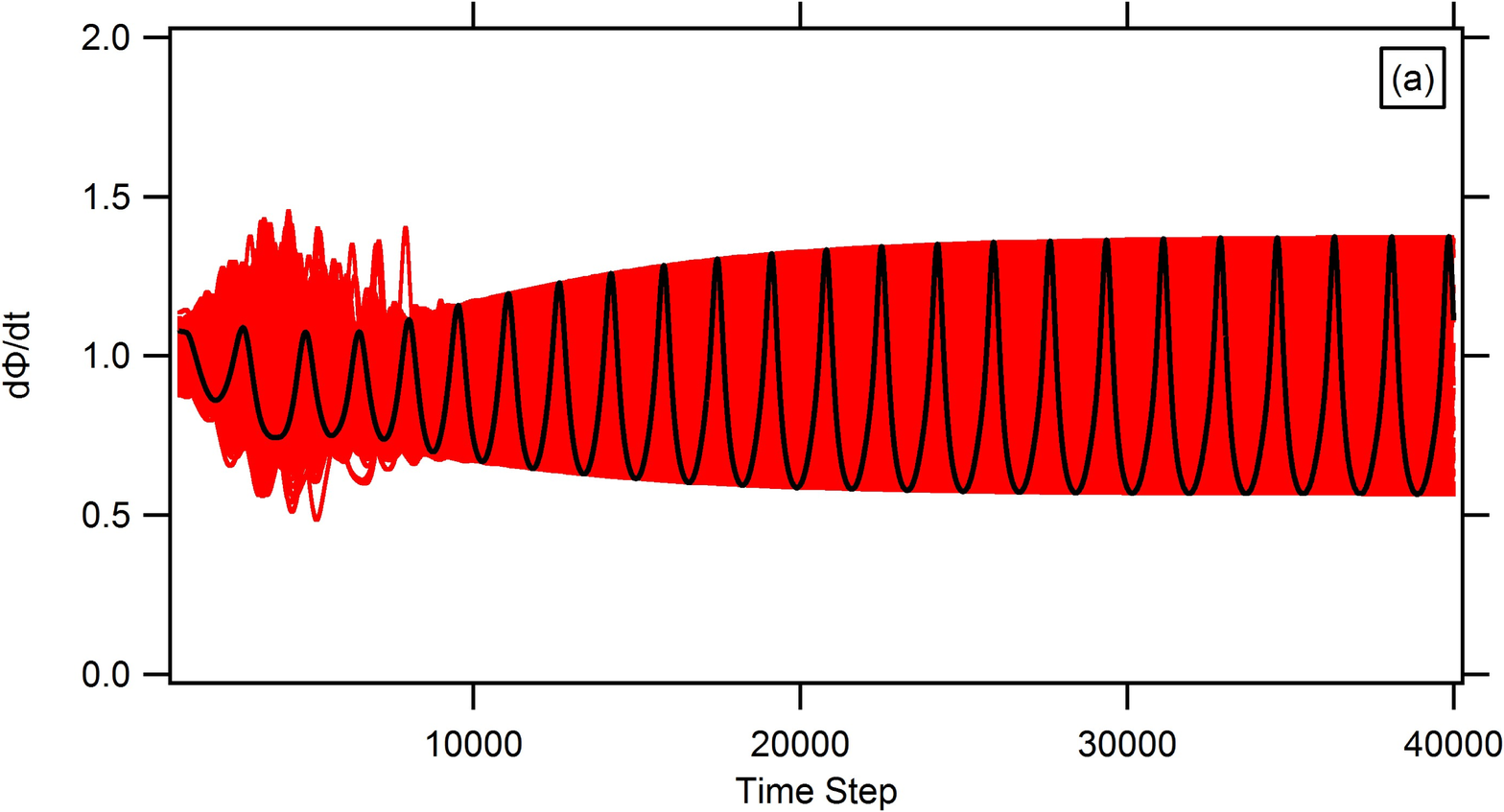}
\caption{The (a) long-time phase pattern and (b) frequency evolution of the system out to 40000 time-steps, for $T=6$ and $\epsilon=0$. All frequencies are included in the lighter (red) traces while a single oscillator's frequency is highlighted with the darker trace (black). Notice that the frequencies fall into a periodic oscillation.}
\label{patterns}
\end{figure}

Further, these patterns face effects that do not have analogues in the 1D setting. First, the second dimension will allow many more ``degenerate'' patterns. That is, some patterns are equally suitable accommodations for the time-delay between oscillators. For example, a plane-wave mode traveling along the x-axis is just as effective an accommodation as one traveling along the y-axis. Second, no pattern will now satisfy the same percentage of oscillator-pairs as the simpler modes of the 1D ring. Continuing with our plane-wave example, if the wave is traveling in the positive y-direction of a 2D array, then along a line parallel to the x-axis, all oscillators will be forced into a common phase even though there will be substantial time-delays involved between oscillators along that line. This holds true to varying extent for every  oscillator-pair not aligned with the y-axis (the direction of phase-propagation). An example of a simulation that organizes into a type of plane wave is provided in Fig.\ref{patterns}(a). In this figure, we are using a $20\times20$ array of oscillators with $T=6$, $dt=.01$, and $t=40000$. Also, note that the pattern is being displayed such that $2\pi=0$ and $\pi$ is the peak of the oscillation. (This graphical scheme was chosen for reasons that become clear in the case of high $\epsilon$, fast learning.) The plane-wave pattern along the y-axis is evident. This means that the time delay is only being accounted for along the y-axis and so the system is experiencing stronger ``frustration" effects than in the 1D analog. These effects hold true to some extent for every spatial pattern available to the system at significant $T$ values. 

Third, the behaviors of the system when in a static network ($\epsilon=0$) are significantly different from the analogous 1D cases. Returning to our plane wave example in Fig. \ref{patterns}(a), while this pattern shows a clear stability and propagates through the system indefinitely, {\it it no longer fits with our definition of global sync.} That is, the frequencies of all the oscillators do not converge to a single value of $\Omega$. Instead, we find that the oscillators' frequencies themselves reach a stable semi-sinusoidal oscillation as seen in Fig.\ref{patterns}(b) in which the darker trace is the frequency of a single oscillator. This result invalidates the assumptions required by both analytical arguments and the order parameters employed in 1D. These frequency oscillations seen in 2D but not in 1D are very likely due to the frustration effects discussed above, and this is substantiated further when we turn to the case of fast learning.
\begin{figure}
\includegraphics[width=3.0in]{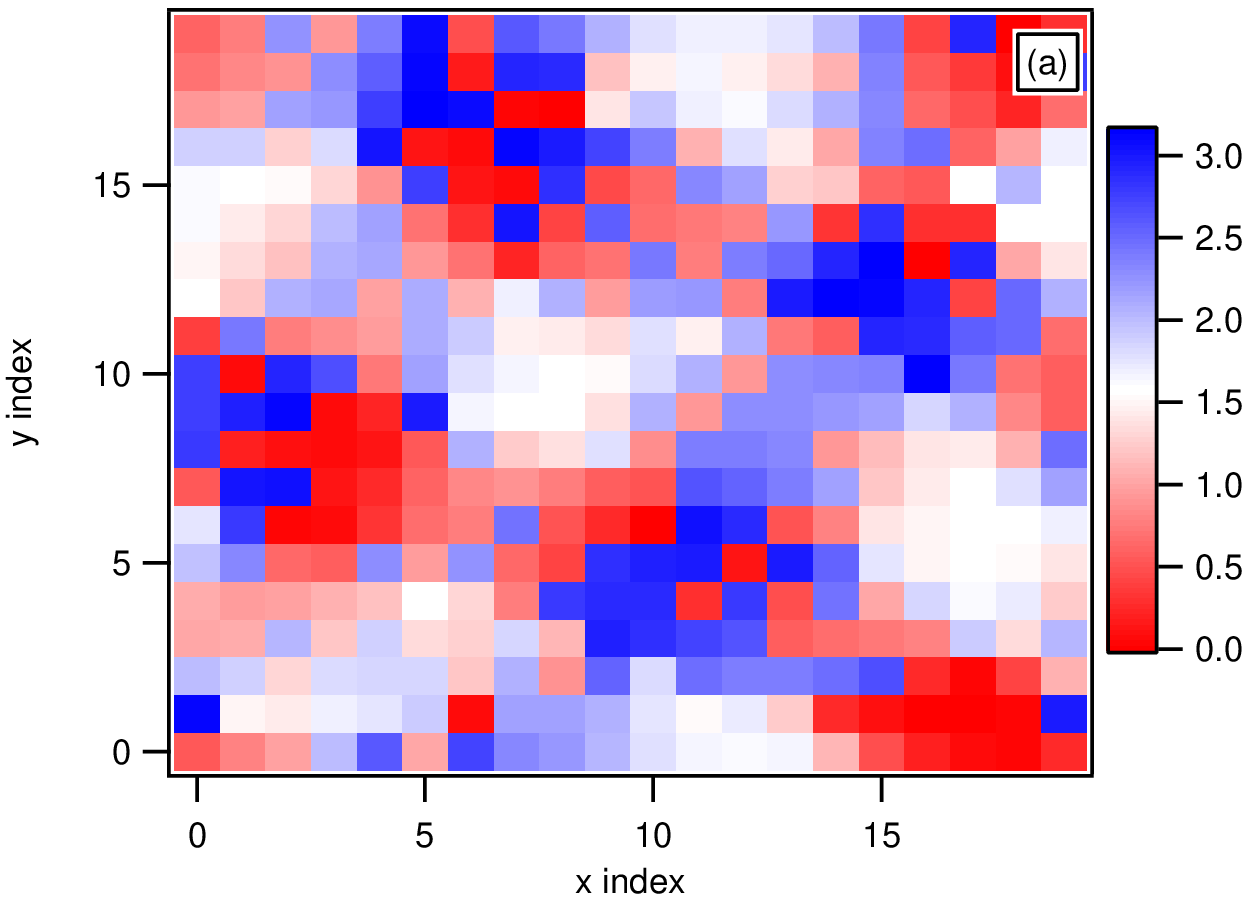}
\includegraphics[width=3.0in]{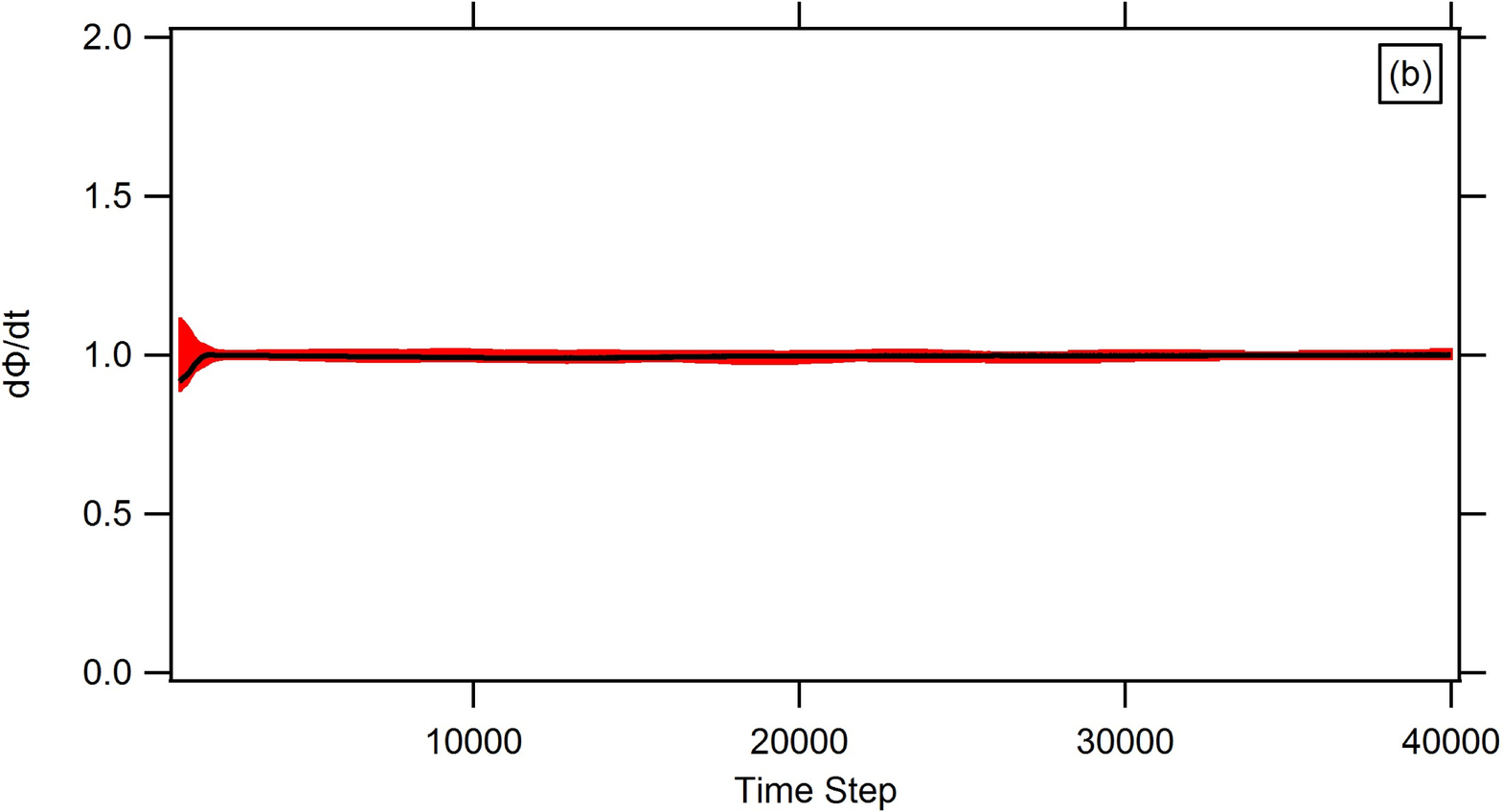}
\caption{The (a) long-time phase pattern and (b) frequency evolution of the system out to 40000 time-steps, for $T=6$ and $\epsilon=0.1$. All frequencies are included in the lighter (red) traces while a single oscillator's frequency is highlighted with the darker trace (black). Now the oscillators are able to stay together in frequency.}
\label{patterns2}
\end{figure}

When the learning-rate parameter is slightly increased from zero (to $\epsilon=0.01$), the situation significantly changes, as summarized by Fig. \ref{patterns2}. The spatio-temporal pattern is characterized by more localized regions that appear to swirl around each other. Furthermore, we observe a concommitant reassertion of frequency synchronization - the frequencies of the oscillators organized in such a pattern are now able to converge. To understand how this convergence is possible in the presence of a nonzero learning rate, let us examine the coupling strengths, $K$. After all, when $\epsilon$ is set to zero, these are held fixed at their initial values of 1, whereas when some learning is turned on, they may adjust themselves organically. 

\begin{figure}
\begin{center}
\includegraphics[width=3.0in]{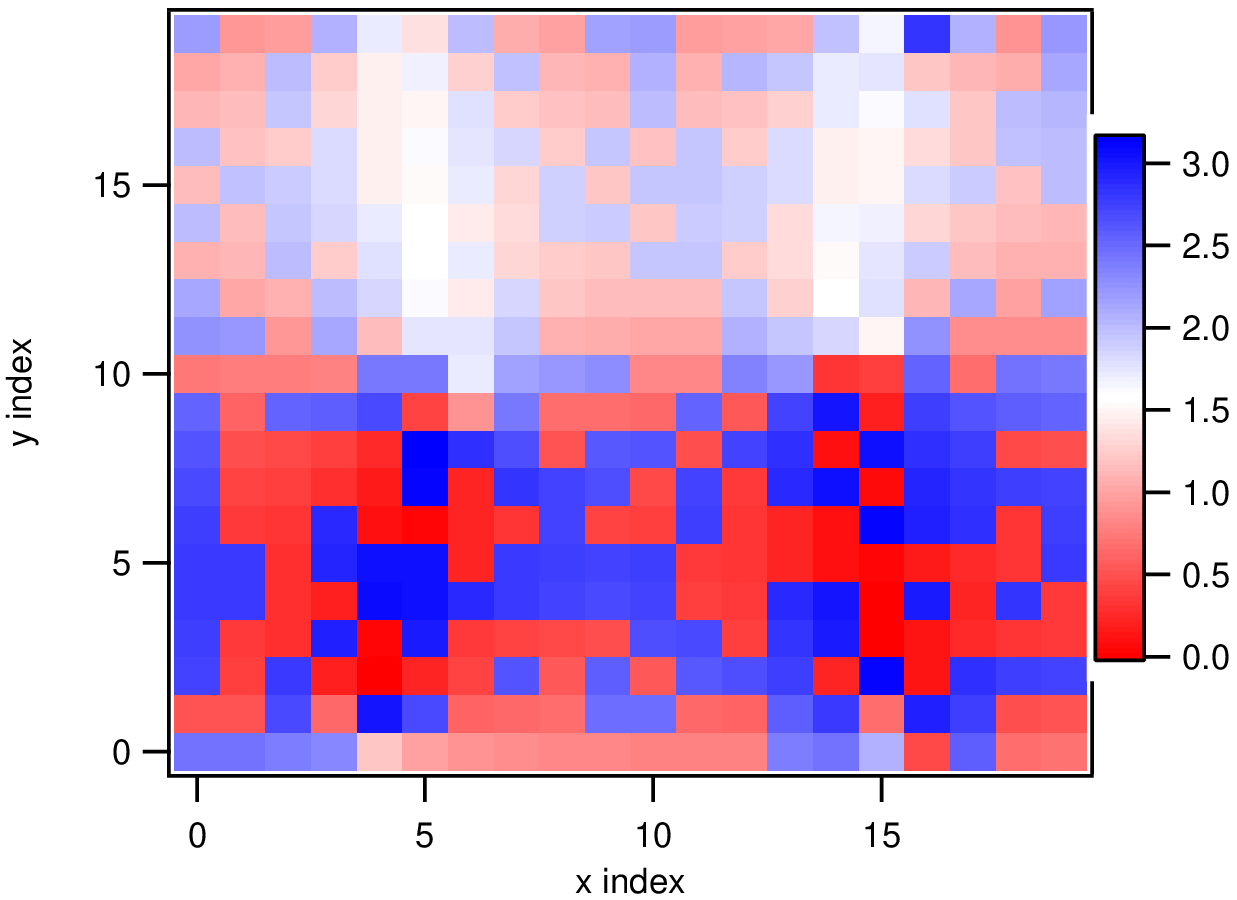}
\includegraphics[width=2.5in]{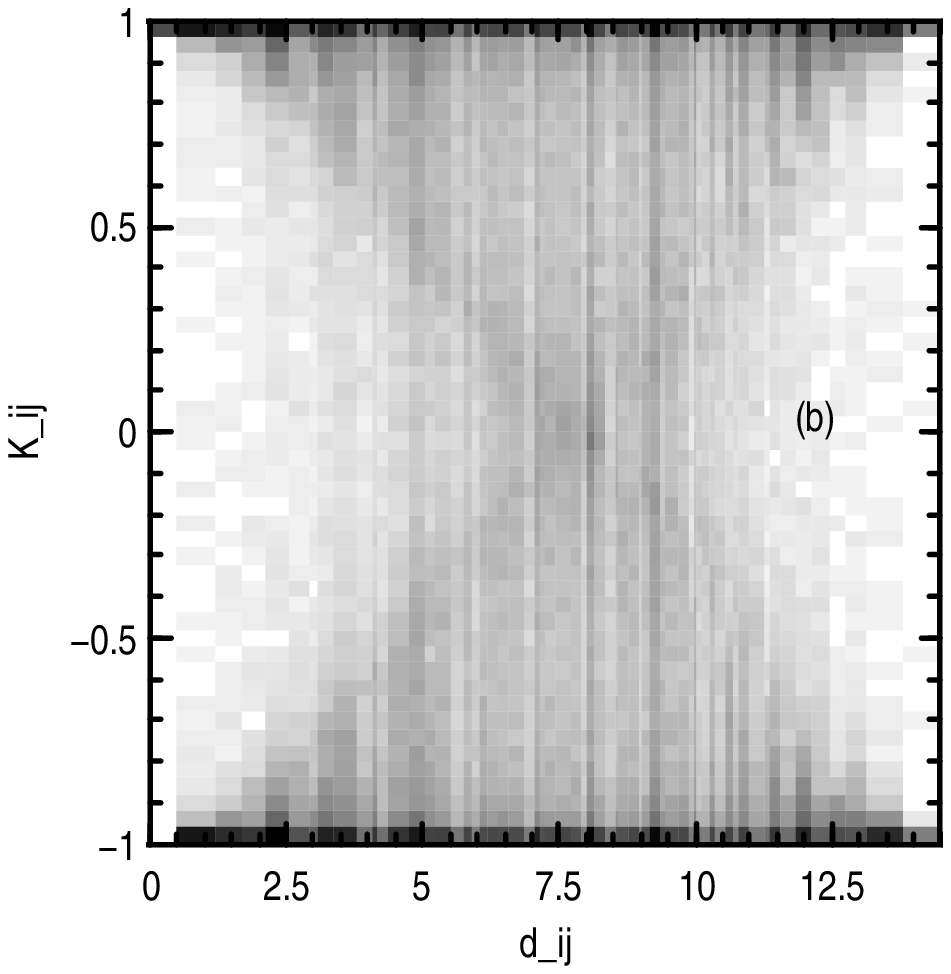}
\caption{$T$=2, $\epsilon=0.1$, (a) The shading (color) indicates the phase of each oscillator. A quasi-1D mode along the x-axis can be discerned. (b) The coupling strengths of all pairs as a function of their distance. The gray-scale indicates the relative number density of oscillator-pairs at the given coupling strength and distance. It is evident that K $\pm 1$ is still realized by many oscillator-pairs regardless of their relative distance, but a large fraction of pairs smear out over much of this plane. The only low-density areas correspond to small distances and low coupling strengths.}
\label{2D}
\end{center}
\end{figure}

Figure \ref{2D} shows the situation for $\epsilon=0.01$ and $T=2$. The time-delay parameter, $T$, is low enough that a planar front in the y-direction can emerge even in the presence of learning, as evidenced by the spatial phase distribution in Fig. \ref{2D}(a).  
The effect of frustration on the network coupling structure is captured by Fig. \ref{2D}(b), which is similar to Fig. \ref{Kvd}, but now for a 2D lattice. Although the phase pattern resembles a plane-wave in the y-direction, the system's $K_{ij,kl}$ organization is not quasi-1D: even though we still see increased point-densities at $K= \pm 1$, these regions now encompass a far smaller percentage of all points. The remaining values do not fall on a well-defined curve (as before in 1D) but are broadly dispersed. This makes sense in terms of the 2D geometry. The darker lines in the grayscale of Fig. \ref{2D}(b) are fairly well matched by the function $K_{ij}=\pm \alpha \cos(\Omega*2T/N*d_{ij})$, with $\Omega=1$ (approximately true for all oscillators), which is the curve we would get for pairs oriented against the mode in the y-direction, and which is what we would get in 1D. Since this represents the fastest that oscillator-pairs can come out-of-phase with distance, we would not expect any dots left of those lines (before they cross); indeed, the density of points is relatively lower in this region. 

However, it is worth noting how many of the coupling strengths are no longer at $\pm1$ and it appears to be this behavior that underlies the reemergence of the stable $\Omega$, as demonstrated by Fig.\ref{patterns2}(b). That is, when $\epsilon$ is nonzero, the oscillators are able to decrease the strength of their interaction and even entirely decouple, $K_{ij,kl}\rightarrow 0$, from those peers that cause the most frustration. Finally, the patterns exhibited by the fast learning case maybe much more complex spatially than those that emerge from the no learning case for the same delay. We can observe this in Figs.\ref{patterns}(a) and \ref{patterns2}(a), where $T=6$ for both panels, yet the fast learning case always organizes into spiraling clusters of oscillators exactly $\pi$ out-of-phase, and the no-learning case forms a simple plane-wave every time.

\section{Conclusion}
\label{sec:conclusion}

We have investigated 1D and 2D arrays of time-delayed, dynamically-coupled Kuramoto oscillators. This combination model generalizes the original Kuramoto model in two ways that promise greater applicability to neural network dynamics; neither separate variation (time delay or dynamics coupling) as treated previously in the literature seems adequate in the neurophysiological context. Here we showed that the combination of these two modifications gives rise to novel synchronization phenomena that are absent in models with only one of them. In particular, in 1D we postulated novel two-cluster integer and half-integer mode patterns and numerically observed their selection in parameter space. We also derived analytically the frequency of synchronization associated with all of these 1D modes and obtained excellent numerical agreement. In order to ascertain the role of dimensionality, we explored the dynamics of this model on 2D arrays of oscillators. We found that enhanced frustration in 2D significantly changes the behavior of the combination model relative to the 1D case. Specifically, with static coupling (no learning) the system is unable to establish a fixed synchronization frequency. In the presence of dynamic coupling, coherent spatio-temporal structures emerge characterized by a single synchronization frequency. In these cases, the network coupling structure adjusts itself so as to reduce phase frustration between oscillators.

\acknowledgments
L.Q.E. acknowledges the helpful discussion with Dr. David Mertens, as well as his assistance with one figure.

\end{document}